\documentclass[12pt]{article}
\usepackage[cp1251]{inputenc}
\usepackage[english]{babel}
\usepackage{epsfig}
\usepackage{amssymb}
\usepackage{amsmath}
 0
\begin{document}

 \title{Quantum corrections to classical solutions via generalized zeta-function }
\author{Anatolij Zaitsev*, Sergey Leble
\\  Gdansk University of Technology, \small \\ Faculty of Applied
Physics and Mathematics, \\\small ul. Narutowicza 11/12, 80-952
Gdansk, Poland,\\ *Kaliningrad State University, \small Faculty of
Physics,
\\ \small Al.Nevsky st. 14, 236041,  \small Kaliningrad, Russia.
 \\\small leble@mif.pg.gda.pl \\  \\[2ex] }
\maketitle

\renewcommand{\abstractname}{\small }

 \begin{abstract}
 A general algebraic method of quantum corrections evaluations is
 presented.
Quantum corrections to a few classical solutions of
Landau-Ginzburg model (phi-in-quadro) are calculated in arbitrary
dimensions.
 The Green function for heat equation with soliton potential is constructed by Darboux transformation.  The
generalized zeta-function is used to evaluate the functional
integral and  corrections  to mass in quasiclassical
approximation. Some natural generalizations for matrix equations
 are discussed in conclusion.
\end{abstract}

\section{Introduction}

In the papers of V.Konoplich \cite{Ko} quantum corrections to a
few classical solutions by means of Riemann zeta-function are
calculated in  dimensions $d > 1$. Most interesting of them are
the corrections to the kink - the separatrix solution of field
$\phi^4$ model. The method of \cite{Ko} is rather complicated and
it is desired to simplify it, that is the main target of this note
with some nontrivial details missed in \cite{Ko}. The suggested
approach open new possibilities; for example it allows to show the
way to calculate the quantum corrections to matrix models of
similar structure, Q-balls \cite{CE} and periodic solutions of the
models. The last problem is posed in the review \cite{TD}.

We mainly concentrate on a technique of evaluation of the quantum
corrections to static one-dimensional solutions of the
d-dimensional Landau-Ginzburg model. Starting from a sketch of the
method, in this introduction, we also give the detailed
description of the solutions. The zeta-function is introduced in
the next section, the last section is devoted to the evaluation of
the diagonal of the Schrodinger operator kernel.

The nonlinear Klein-Gordon equation in the case of static
one-dimensional solutions is reduced to
\begin{equation}\label{KG}
    \phi''- V'(\phi)=0, \phi = \phi(x), x\in R.
\end{equation}
Suppose the potential $V(\phi)$ is twice continuously
differentiable; it guarantees existence and uniqueness of the
equations' (\ref{KG}) Cauchy problem solution. The first integral
of (\ref{KG}) is given by
\begin{equation}\label{E}
    E = \frac{1}{2}(\phi')^2-V(\phi),
\end{equation}
where E is the integration constant. The equation (\ref{E}) is
ordinary first-order differential equation with separated
variables. As the phase method shows, the solutions of this
equations belong to the following families: constant, periodic,
separatrix and the so-called "passing" one \cite{TD}.

The approximate quantum corrections to the solutions of the
equation (\ref{KG}) are obtained if the Feynmann functional
integral by trajectories is evaluated by the continual stationary
phase method. It gives the following relation
\begin{equation}\label{S}
    \exp[-\frac{S_{qu}}{\hbar}]\simeq\frac{A}{\sqrt{D}},
\end{equation}
where $S_{qu}$ denotes quantum action, corresponding the potential
$V(\phi)$, $A$ - some quantity determined by the vacuum state at
$V(\phi) = 0$, and $\det D$ is the determinant of the operator
\begin{equation}\label{D}
   D = - \partial^2_x - \Delta_y + V''(\phi(x)).
\end{equation}
The variable $y \in R^{d-1}$ stands for the transverse variables
on which the solution $\phi(x)$ does not depend. The operator D
appears while the evaluation of the second variational derivative
of the quantum action functional (which enter the Feynmann
trajectory integral) is provided. For the vacuum action $S_{vac}$
the  relation of the form (\ref{S}) is valid if $S_{qu}$ is
changed to $S_{vac}$ and $D$ is placed by the "vacuum state"
operator $D_0 = -
\partial^2_x - \Delta_y$. Then, the quantum correction
\begin{equation}\label{qucor}
   \Delta S_{qu} =  S_{qu} - S_{vac},
\end{equation}
is obtained by the mentioned twice use of the formula (\ref{S}) as
\begin{equation}\label{qucor1}
   \Delta S_{qu} = \frac{
   \hbar}{2}\ln(\frac{\det{D}}{\det{D_{0}}}).
\end{equation}
Hence, the problem of determination of the quantum correction is
reduced to one of evaluation of the determinants od $D$ and
$D_{0}$. The methodic of the evaluation will be presented in the
following section.

\medskip
\section{The generalized Riemann zeta-function,  preliminaries.}
\medskip
The generalized zeta-function appears in many problems of quantum
mechanics and quantum field theories which use the Lagrangian
$L=({\bf\partial}\phi)^2/2
 - V(\phi)$
and it is necessary to calculate a Feynmann functional integral in
quasiclassical approximation.

The scheme is following. Let $\{\lambda_n\} =  S$ be a set of all
eigenvalues of a linear operator $L$, then, logarithm of the
operator determinant is represented by the formal sum by this set
\begin{equation}\label{lndet}
    \ln(\det L)= \sum_{\lambda_n\in S}\ln\lambda_n.
\end{equation}
Let us next define the generalized Riemann zeta-function
$\zeta_{L}(s)$ by the equality
\begin{equation}\label{zeta}
   \zeta_{L}(s) = \sum_{\lambda_n\in S}\lambda_n^{-s}.
\end{equation}
This definition should be interpreted as analytic continuation
from the half plane $Re s > \sigma $ in which the sum  in
(\ref{zeta}) converge. Differentiating the relation (\ref{zeta})
with respect to $s$ at the point $s=0$ yields
\begin{equation}\label{lndet}
    \ln(\det L)= \zeta_{L}'(0).
\end{equation}
The generalized function (\ref{zeta}) admits the representation
via the diagonal $g_D$ of a Green function of the operator
$\partial_t+L$. The representation is obtained as follows.

Let $\textbf{r} \in  \mathbb{R^d}$ be the set of independent
variables of the operator $L$; particularly, the operator $D$ of
(\ref{D}) depends on $\textbf{r}=(x,\textbf{y}) \in \mathbb{R}
\times \mathbb{R}^{d-1}$, then
 \begin{equation}\label{g}
    (\partial_t+L)g(t,\bf{r},\bf{r_0})=\delta(\bf{r}-\bf{r_0})
\end{equation}
and $$ g(t,{\bf{r},\bf{r_0}})=0, \quad t<0. $$
 There is a representation
in terms of the formal sun
\begin{equation}\label{gD}
    g_D(t,\bf{r},\bf{r_0}) =\sum_n  \exp[-\lambda_n t] \psi_n(\bf r)\psi_n^*(\bf
    r_0),
\end{equation}
where the normalized eigenfunctions $\psi_n(\bf r)$ correspond to
eigenvalues $\lambda_n$ of the operator $D$. Let us introduce the
function
\begin{equation}\label{gamma}
    \gamma_D(t) = \int d {\bf r} g_D(t,\bf{r},\bf{r}) = \sum_n  \exp[-\lambda_n t],
\end{equation}
that follows from (\ref{gD}) and normalization. The Mellin
transformation of (\ref{gamma}) yields in
\begin{equation}\label{gammazeta}
   \zeta_D(s) = \frac{1}{\Gamma(s)}\int_0^{+\infty}t^{s-1}\gamma_D(t)dt
   ,
\end{equation}
where the $\Gamma(s)$ is the Euler Gamma-function.

The generalized zeta-function, defined by the relations
(\ref{gamma},\ref{gammazeta}),   will be referred as the
zeta-function of the operator $D$.

From the relation (\ref{gamma}) for the function $\gamma_D(t)$ it
follows most important property of multiplicity: {\it if the
operator D is a sum of two differential  operators $D = D_1 +
D_2$, which depend on different variables, the following equality
holds}
\begin{equation}\label{gg}
\gamma_D(t)=\gamma_{D_1}(t)\gamma_{D_2}(t).
\end{equation}
We will need the value of the function $\gamma_D(t)$ for the
vacuum state, when the operator $D=D_0 = -\Delta$ is equal to the
d-dimensional Laplacian. In this case the formal sum in the r.h.s
of (\ref{gamma}) goes to d-dimensional Poisson integral
\begin{equation}\label{P}
   \gamma_{D_{0}}(t) = \frac{1}{(2\pi)^d}\int_{\mathbb{R^d}}d{\bf k}
   \exp(-|{\bf k}|^2t) = (4\pi t)^{-d/2}.
\end{equation}

The basic relation (\ref{qucor1}) points to a necessity of
evaluation of the determinants of the operators
\begin{equation}\label{det}
    D=D_0+u(x), \quad D_0=-\partial_x^2-\Delta_y+\lambda
\end{equation}
where $\lambda$ is a positive number and $ x\in R$ is one of
variables, while $\textbf{y}\in R_{d-1}$ is a set of other
variables. The operator $ \Delta_y $ is the Laplacian in d-1
dimensions, u(x) is one-dimensional potential that is defined by
the condition
\begin{equation}\label{pot}
    V''(\phi_0(x)) = \lambda+u(x),
\end{equation}
where $\phi_0(x)$ is the classical static solution of the equation
of motion.

A quantum correction to the action in one-loop approximation for the
classical solution $\phi_(x)$ is calculated via zeta-function by the formula
\begin{equation}\label{qucorM }
    \Delta\epsilon = -\zeta'_D(0)/2,
\end{equation}
where
\begin{equation}\label{zetaM}
    \zeta_D(s) = M^{2s}\int_0^\infty \gamma(t)t^{s-1}dt/\Gamma(s);
\end{equation}
here $\Gamma(s)$ is the Euler gamma function and M is a mass
scale.

The function $\gamma(t)$ in the Mellin integral (\ref{zetaM}) is
expressed via the Green functions difference
 $G(x,y,t;x_0,y_0,t_0)$ and
   $G_0(x,y,t;x_0,y_0,t_0)$
 of $\partial_t+D$  and  $ \partial_t + D_0$ in the
following way; let
$$g(x,t) =  G(x,y,t;x_0,y_0,0) - G_0(x,y,t;x_0,y_0,0),$$
due to the translational invariance along y of operators D and
$D_0$ the function g does not depend on y; the contraction of
$G_0$ is necessary for deleting of ultraviolet divergence. Then
the following formula is valid
\begin{equation}\label{gam}
    \gamma(t) = \int_{-\infty}^\infty g(x,t)dx.
\end{equation}

In the case of static solutions of the $\varphi^4$ model the
potential and the Lagrangian are determined by the formulas
$$V(\varphi) = g\varphi^4/4 - m^2\varphi^2/2,$$\qquad
$$L =
-(\varphi')^2 - g\varphi^4/4 + m^2\varphi^2/2,$$
 therefore the
equation of motion has the form
\begin{equation}\label{eqmot}
    \varphi''(x) + m^2\varphi - g\varphi^3 = 0.
\end{equation}
Its separatrix solution is the kink
\begin{equation}\label{kink}
    \varphi_0 = m\tanh(mx/\sqrt{2})/\sqrt{g}.
\end{equation}

After the substitution of (\ref{kink}) into (\ref{pot}) we obtain
the following potential u(x) :
\begin{equation}\label{pokink}
    u(x) = -6b^2/ch^2(bx),
\end{equation}
with the meaning of the constant $  b = m/\sqrt{2}. $ As a result
the two-level reflectionless potential of one-dimensional
Schr{\accent "7F o}dinger equation $ -
\partial^2_x + u(x) $ appears. Eigenvalues  and the
normalized eigenfunctions of which are correspondingly (its
numeration is chosen from above to lowercase).
$$\lambda_1 = - b^2, \qquad \psi_1(x) = \sqrt{3b/2} \sinh(bx)/cosh^2(bx);$$
$$\lambda_2 = -4b^2,\qquad \psi_2(x) = \sqrt{3b}/2\cosh(bx)$$
In the next section (sec.3) we suppose for generality that the
potential u(x) from (\ref{pot}) has the form of n-level
reflectionless potential
\begin{equation}\label{nlevel}
    u(x) = -n(n+1)b^2/\cosh^2(bx).
\end{equation}
with  eigenvalues $\lambda_m = -m^2b^2$, $m=1,...,n$. We would
note for further generalization that this function may be
considered as degenerate limit of n-gap Lame potential of Hill
equation (see conclusion). Such potentials correspond to higher
solitonic models.

The kink case corresponds $n = 2, \lambda = n^2b^2$. The quantum
correction to its action will be calculated in sec.4 from general
formulas of sec.3.


\medskip
\section{Zeta-function representation and its derivative at zero
point.}
\medskip
Formula (\ref{gam}) for $\gamma(t)$ may be simplified if one
expresses the integrand via the Green function of one-dimensional
Schr{\accent "7F o}dinger operator with nonzero and zero
potentials.

{\bf Proposition 1}.  {\it Let  $G^{(1)}(x,t;x_0,t_0)$ and
   $G_0^{(1)}(x,t;x_0,t_0)$ are Green functions of one dimensional evolution
operators
$\partial - \partial_x^2$
 + u(x) and $\partial - \partial_x^2$. And let
$$ e(x,t) = G^{(1)}(x,y,t;x_0,y_0,t_0) - G_0^{(1)}(x,y,t;x_0,y_0,t_0),$$
then the following representation for $g(x,t)$ from (\ref{gam}) is
valid:
\begin{equation}\label{gt}
    g(x,t) = (4\pi t)^{(1-d)/2} exp[-\lambda t] e(x,t).
\end{equation}}
{\bf Proof}. The Green functions $G(x,y,t;x_0,y_0,t_0)$ and
   $G_0(x,y,t;x_0,y_0,t_0)$ are easily expressed via  $G^{(1)}(x, t;
x_0, t_0)$ and
   $G_0^{(1)}(x, t;x_0, t_0)$ by means of Fourier transformations by \textbf{y};
putting in these expressions $x_0 = x, \mathbf{y_0} = \mathbf{y},
t_0 = t$ we get
$$g(x,t) = exp[-\lambda t] e(x,t)\int_{\mathbb{R}^{d-1}} exp[-\mid \mathbf{k}\mid^2t]d\mathbf{k}/(2\pi)^
{d-1}$$ (here \textbf{k} is the vector Fourier transform
parameter). The known Poisson integral ()\ref{P} of power d-1
appears; using its value one goes to (\ref{gt}).

{\bf Corollary}. The following trace formulae takes place:
\begin{equation}\label{tr}
    \gamma(t) = (4\pi t)^{(1-d)/2} exp[-\lambda t] \gamma_0(t),
\end{equation}
  where
$$\gamma_0(t) = \int_{-\infty}^{\infty} e(x,t)dx$$
\textbf{Remark 1}. Formula (\ref{tr}) is named as trace formula
because $\gamma_0(t)$ is the difference of traces of reciprocals
of operators $\partial - \partial_x^2
 + u(x)$ and $\partial - \partial_x^2$. This sense of $\gamma_0(t)$ is reserved
for possible generalizations of the problem under consideration
(see conclusion).

Now we go to the determination of the explicit form of the
function $\gamma_0(t)$. From the results of \cite{LeZa3} it is
possible to extract the following

{\bf Proposition 2}. {\it Let the integral
\begin{equation}\label{13}
    \hat {e}(x,t) = \int_0^{\infty}exp[-p t]dt
\end{equation}
be the Laplace transformation of e(x,t), then
\begin{equation}\label{14}
    \hat{e}(x,p) =\frac{1}{\sqrt{p}} \sum_{\nu=1}^{n}\psi_{\nu}^2(x)\nu b/(p -
\nu ^2b^2),
\end{equation}
where $\psi_\nu(x)$ is a normalized eigen function that correspond
to the eigenvalue $\lambda_\nu = - \nu^2b^2, \nu = 1,..,n,$ of the
one- dimensional operator with the potential (\ref{nlevel}).}

From (\ref{tr})-(\ref{14}) we obtain   the Laplace transform of
the function $\gamma_0(t)$
$$\int_{-\infty}^{\infty}\hat{e}(x,p)dx = \frac{1}{\sqrt{p}}\sum_{\nu=1}^n \nu b/(p - \nu^2b^2),$$
and by the table of inverse Laplace transformations \cite{BE} one
finds
\begin{equation}\label{g0}
    \gamma_0(t) = \sum_{\nu=1}^n exp[\nu^2 b^2t] Erf(\nu b
\sqrt{t}).
\end{equation}

 Now one can derive the basic result
(probably absent in publications) of the section.

{\bf Theorem.} {\it Zeta-function $\zeta_D(s)$ has the following
representations:}
\begin{equation}\label{zetaD1}
    \zeta_D(s) = \frac{4 \Gamma(s + 1 - d/2)}{(4\pi)^{d/2}\Gamma(s)}\sum_{\nu=1}^n(\nu b)^{d-1}(\frac{M}{\nu b})^{2s}
\int_0^1(\tau^2 - 1 + \frac{\lambda }{\nu^2 b^2})^{d/2-1-s}d\tau;
\end{equation}
 \begin{equation}\label{zetaD2}
    \zeta_D(s) = b(\frac{\lambda}{4\pi})^{\frac{D}{2}-1}(\frac{M^2}{\lambda})^s\frac{\Gamma(s+1-1/2)}{\pi
\Gamma(s)} \sum_{\nu=1}^{n}\nu
F(s+1-d/2,1,3/2;\frac{\nu^2b^2}{\lambda}).
\end{equation}

 {\bf Proof.} The integral that define error
function may be transformed by the simple change of variables to
the form
$$ Erf(z) = 2z\int_0^1exp[-z^2\tau^2]d\tau/\sqrt{\pi}.$$
Substituting this representation to the formulae (\ref{g0}) and
further in (\ref{tr}), (\ref{gam}) and (\ref{zetaM}), by means of
analytical prolongation in s, one arrives at the expression
(\ref{zetaD1}). The integral in it is transformed via
Hypergeometrical function \cite{Ku} that gives (\ref{zetaD2}).
This expression is cited from esthetic point; further only the
formulae (\ref{zetaD1}) is used.
\section{Quantum correction to the kink and periodic solutions mass in the   1,2,3,4 dimensions.}
Now we go to the  zeta-function derivative evaluation. The meaning
of $\zeta_D'(0)$ should be calculated in different ways for d=1,
even $d > 2$, and odd ones. For the beginning we formulate and
prove the following useful intermediate result.

{\bf Proposition 3}. {\it When d=1, $\lambda > n^2 b^2$ and when d
= 2N-1, N = 2,3,.. , $\lambda \geq n^2b^2$ the following equality
is valid
\begin{equation}\label{18}
    \zeta_D'(0) = 4(4\pi)^{1/2 - N} \Gamma(3/2 - N)\sum_{\nu=1}^n(\nu b)^{2(N-1)}
R_N(\lambda/\nu^2b^2),
\end{equation}
where,
\begin{equation}\label{19}
    R_N(z) = \int(\tau^2 - 1 +z)^{N-3/2}d\tau.
\end{equation}
When d = 1, $\lambda\geq n^2b^2,$
\begin{equation}\label{20}
    \zeta_D'(0) = \frac{2}{\sqrt{\pi}}2\sum_{\nu=1}^{n-1}R_1(n^2/\nu^2) - \frac{d}{ds}(\frac{\Gamma(s +
1/2)}{\Gamma(s + 1)}(\frac{M}{nb})^{2s})|_{s=0}.
\end{equation}
When d = 2N, N = 1,2,.., $\lambda \geq n^2b^2$,
\begin{equation}\label{21}
    \zeta_D(0) = \frac{4(-1)^{N+1}}{(4\pi)^{N}(N-1)!}\sum_{\nu=1}^n (\nu b)^{2N-1}((\gamma_N +
2\ln[\frac{M}{\gamma b}])P_{N-1}(\frac{\lambda}{\nu^2b^2}) -
 J_{N-1}(\frac{\lambda}{\nu^2b^2}),
\end{equation}
where. for N = 0,1,2,...,
\begin{equation}\label{22}
    P_N(z) = \int_0^1(\tau^2 - 1 + z)^Nd\tau,\quad J_N(z) = \int_0^1(\tau^2 - 1 + z)^N
\ln(\tau^2 -1 + z)d\tau,
\end{equation}
and
 $$\gamma_N =
\sum_{j=1}^{N-1} 1/j.$$}

{\bf Proof. } At d =2N1 the formulae (\ref{zetaD1}) converts in
the following one:
\begin{equation}\label{23}
    \zeta_D(s) = 4(4\pi)^{1/2-N}\frac{\Gamma(s + 3/2 - N)}{\Gamma(s)}\sum_{\nu=1}^n(\nu b)^{2(N-1)}
(\frac{M}{\nu b})^s\int_0^1(\tau^2-
 1 + \frac{\lambda}{\nu^2 b^2})^{N-s-3/2}d\tau
\end{equation}
In the cases $N \geq 2, \lambda \geq n^2b^2,$ and $N = 1, \lambda
> n^2b^2$, the function in the right-hand side is analytical in
the vicinity of the point $s = 0$ and
$\lim_{s\rightarrow0}\zeta_D(s) = 0$. Therefore
 $\zeta'_D(0) = \lim_{s\rightarrow0}\zeta_D(s)/s$.
As $\lim_{s \rightarrow 0}\frac{1}{s\zeta_D(s)}= 1$, from
(\ref{23}) one immediately obtain (\ref{18}).

The case $d=1$, $\lambda = n^2b^2$ is singular because the
integral in the last term of (\ref{23}) diverges at s = 0 (when
$\nu = n$ it degenerates into the integral $\int_0^1\tau^{ -1 -
2s}d\tau )$. However, due to $1/\Gamma(s) \simeq s$ when $ s
\rightarrow 0$, the term with $\nu = n$ is continued analytically
till $s = 0$ but now $ \lim_{s \rightarrow 0}\zeta_D(s) \neq 0$.
Really, evaluating the integral in this term inside the region of
convergence $\Re s < $,, one gets $\int_0^1\tau^{-1-2s}d\tau =
-1/2s$. Thus the last term in (\ref{23}) in the case of
d=1,$\lambda = n^2b^2$ yields in analytical (in the vicinity of
$s=0$) function
$\frac{\Gamma(s+1/2)}{\sqrt{\pi}\Gamma(s+1)}(\frac{M}{nb})$.
Separating further in r.h.s. of (\ref{23}) the last term and
calculating derivatives of the rest terms in $s=0$ in the same
manner, one arrives at \eqref{20}.

 In even dimensions $d=2N$, the
formula (\ref{zetaD1}) takes the form
\begin{equation}\label{zeteven}
    \zeta_D(s) = \frac{4 \Gamma(s + 1 - N)}{(4\pi)^{N}\Gamma(s)}\sum_{\nu=1}^n(\nu b)^{2N-1}(\frac{M}{\nu b})^{2s}
\int_0^1(\tau^2 - 1 + \frac{\lambda }{\nu^2 b^2})^{N-1-s}d\tau;
\end{equation}
Differentiating it in $s=0$ with account of the equality
$$\frac{d}{ds}\ln\frac{ \Gamma(s + 1 - N)}{\Gamma(s)}|_{s=0} = \gamma_N$$
one obtains \eqref{21}. The statement 3 gives an intermediate
result; the final form derivation needs the calculation of
integrals in (\ref{19},\ref{22}).

Let us start with
$$R_1(z)=\int(\tau^2 - 1 +z)^{-1/2}d\tau = \frac{1}{2} \ln\frac{\sqrt{z}+1}{\sqrt{z}-1}, z>1.$$
Therefore $ \zeta_D(0) = \ln\prod_{\nu=1}^n\frac{\sqrt{z}+\nu
b}{\sqrt{z}-\nu b} $, d=1, $\lambda > n^2b^2.$

It is easy to check (see e.g. \cite{Ku})
$$\frac{d}{ds}\ln\frac{ \Gamma(s + 1/2)}{\Gamma(s+1)}|_{s=0}  = -\ln 4;$$ using this one finds from
(\ref{20}) for d=1, $\lambda=n^2b^2$:
$$\zeta_D(0) = \ln (\frac{4C_{2n-1}^nn^2b^2}{M^2}). $$

The values of $\zeta_D(0)$ are calculated by the recurrences
$$
R_{N+1}(z) = \frac{2N-1}{N}(z-1)R_{N}+\frac{z^{N-1/2}}{2N}, \quad
N=1,2,...,
$$
$$
P_{N+1}(z) = \frac{2(N+1)}{2N+3}(z-1)P_{N}+\frac{z^{N+1}}{2N+3},
\quad N=0,1,2,...,
$$
$$
J_{N+1}(z) = \frac{2N-1}{N}(z-1)R_{N}+\frac{1}2N{z^{N-1/2}}, \quad
N=0, 1,2,.. ;
$$
that start from
$$
R_1, \quad P_0 = 1,\quad J_0(z) =
2\sqrt{z-1}\arcsin\frac{1}{\sqrt{z}};
$$
 (obtained by integrating by parts in integrals of
(\ref{19}) and (\ref{22})). After all transformations, the
following expressions for $d=1.2.3.4$ are collected at the next

 {\bf Statement 4}.{\it At d = 1,
$$ \zeta'_D(0) = \ln \prod_{\nu = 1}^n\frac{\sqrt{\lambda}+\nu b}{\sqrt{\lambda}-\nu b} $$
if $\lambda>n^2b^2$; otherwise,
\begin{equation}\label{zeta1}
   \zeta'_D(0) = \ln(\frac{4C_{2n-1}^n n^2b^2}{M^2}),
\end{equation}
if $\lambda = n^2b^2$.

 At d=2
\begin{equation}\label{d2}
\begin{array}{c}
 \zeta'_D(0) = \frac{n(n+1)b}{2\pi} (ln\frac{M^2}{\lambda})-\frac{2}{\pi}\sum_{\nu=1}^{n}\sqrt{(\lambda-\nu^2b^2)}
 \arcsin\frac{\nu b}{\sqrt{\lambda}},\quad \lambda>n^2b^2 \\
 \zeta'_D(0)   = \frac{n(n+1)b}{ \pi} (\ln\frac{M}{nb})-\frac{2b}{\pi}\sum_{\nu=1}^{n}\sqrt{(n^2b^2 - \nu^2)}
 \arcsin\frac{\nu}{n}, \quad \lambda=n^2b^2 \\
\end{array}
\end{equation}
At d=3
\begin{equation}\label{d2}
\begin{array}{c}
 \zeta'_D(0) = - \frac{n(n+1)b\sqrt{\lambda}}{4\pi} + \sum_{\nu=1}^{n} (\lambda-\nu^2b^2)
 \ln\frac{\sqrt{\lambda}+\nu b}{\sqrt{\lambda}-\nu b},\quad \lambda>n^2b^2 \\
 \zeta'_D(0)   = -\frac{b^2}{4\pi}( n^2(n+1)b+  \sum_{\nu=1}^{n} (n^2  - \nu^2)
 \ln\frac{\nu+n}{n-\nu}), \quad \lambda=n^2b^2 \\
\end{array}
\end{equation}
At d=4
\begin{equation}\label{d2}
\begin{array}{c}
 \zeta'_D(0) = \frac{1}{3\pi^2}\{ \frac{n(n+1)b}{8}(n(n+1)b^2-3\lambda)\ln\frac{M^2}{\lambda}\\
  + \frac{n(n+1)b}{8}(n(n+1)b^2-7\lambda) + \sum_{\nu=1}^{n}
  (\lambda-\nu^2b^2)^{3/2}
 \arcsin\frac{\nu b}{\sqrt{\lambda}}\},\quad \lambda>n^2b^2; \\
 \zeta'_D(0)   = - \frac{b^3}{3\pi^2}\{ \frac{n^2(n+1)(13n-8)}{24}\ln\frac{M}{nb}\\
  + \frac{n^2(n+1)(2n-1)}{4}- \sum_{\nu=1}^{n-1}
  (n^2-\nu^2)^{3/2}\arcsin\frac{\nu }{n}\}, \quad \lambda=n^2b^2. \\
\end{array}
\end{equation}}
These formulas (by other method and notations) at $d=2,3,4$ were
derived in \cite{Ko}.

In the case of the kink (see the sec. 2). n=2, b=$m/\sqrt{2},
\lambda=4b^2=2m^2$, basing on the case $\lambda = n^2b^2$, one
obtains the quantum correction to the mass  $\Delta \epsilon =
\zeta'_D(0)$

\begin{eqnarray}
  d=1: \quad\zeta'_D(0)  &=& -2\ln\frac{M}{2\sqrt{6}m}, \quad \\
  d=2: \quad\zeta'_D(0) &=& -\frac{3\sqrt{2}m}{\pi}(1+\ln\frac{M}{\sqrt{2}m}) \\
   d=3: \quad \zeta'_D(0)  &=&  -\frac{3m^2}{8}(\ln3+4)  \\
  d=4: \quad \zeta'_D(0)  &=&
  \frac{m^3}{8}(\frac{1}{4\sqrt{6}}-\frac{3}{2\sqrt{2}\pi}(1+\ln\frac{M}{2\sqrt{6}m}))
\end{eqnarray}
These expressions reproduce the formulas for kink mass correction
from \cite{Ko} for d=2,3,4.

\section{Conclusion}
Let us note that all results related to the scalar nonlinear
Klein-Gordon equation may be applied directly to many-component
model $\phi(x)\in \mathbb{S^m}$ with account of $\mathbb{SO}(m)$
symmetry. The equation (\ref{KG}) takes the form
\begin{equation}\label{KGm}
   - \phi''+ V'(||\phi||)\frac{\phi}{||\phi||} = 0.
\end{equation}
The scalar operator D is defined by (\ref{D}) goes to the matrix
one
\begin{equation}\label{Dm}
   D = [- \partial^2_x - \Delta_y +  V'(||\phi||)\frac{\phi}{||\phi||}]I_m + [(V"(||\phi||)) -
   V'(||\phi||)\frac{\phi\otimes\phi}{||\phi||^3}].
\end{equation}
The technique developed in this paper is transported to quantum
corrections for periodic static solutions of $\phi^4$ model:
\begin{equation}\label{24}
\phi_0(x) =
\frac{km}{1+k^2}\sqrt{\frac{2}{g}}sn(\frac{mx}{\sqrt{1+k^2}};k),\quad
0< k \leq 1;
 \end{equation}
 where $k$ is a module of the elliptic function. When $k=1$ the
 formula (\ref{24}) goes to one for kink (\ref{kink}). The
 substitution of (\ref{24}) into (\ref{pot}) yields in two-gap Lame potential
that is embedded in Darboux Transformations theory by chain
representation \cite{BL} that give a possibility to derive the Green
function analogue for this case. The results will be published
elsewhere. Some recent papers open new field for applications
\cite{B}.


\begin{thebibliography}{bog-kon98-1}
\bibitem{Ko} Konoplich R.V. Quantum corrections calculations to nontrivial
classical solutions via zeta-function.TMP,1987,v73,p 379-392. The
zeta-function method in field theory at finite temperature.
(Russian)
 Teoret. Mat. Fiz. 78 (1989), no. 3, 444--457; translation in Theoret. and Math. Phys. 78 (1989),
  no. 3, 315--325.
\bibitem{CE} Cervero J.M, Estevez P.G. Exact two-dimensional Q-balls near
the kink phase Phys.Lett.B,v176,p139-142,1986.

\bibitem{TD} Tuszy\'nski, J. A.; Dixon, J. M.; Grundland, A. M. Nonlinear field theories and
non-Gaussian fluctuations for near-critical many-body systems.
Fortschr. Phys. 42 (1994), no. 4, 301--337.

\bibitem{LeZa3} Leble S., Zaitsev A, The Modified Resolvent for the One-dimensional Schrodinger Operator
with a reflectionless potential and Green Functions in
Multidimensions  {\it J.Phys. A:Math.Gen.} v.28 (1995)
p.L585-L588.

\bibitem{Suk04} Sukumar, C. V. Green's functions, sum rules and matrix
elements for SUSY partners. J. Phys. A 37 (2004), no. 43,
10287--10295.


\bibitem{NMPZ} Novikov S.P. Manakov S.V. Pitaevski L.P. Zakharov V.E.
Theory of Solitons. Plenum, New York, 1984.

\bibitem{TMCD} Tuszy\'nski, J. A.; Middleton, J.; Christiansen, P. L.; Dixon, J. M. Exact eigenfunctions
of the linear ramp potential in the Gross-Pitaevskii equation for the Bose-Einstein condensate.
Phys. Lett. A 291 (2001), no. 4-5, 220--225.

\bibitem{BE} Tables of Integral Transforms
A Erdelyi, H Bateman - 1954 - McGraw-Hill New York.


\bibitem{Ku} Kuznetsov D.S. Special Functions. Vysshaia Shkola,1965,(in Russian).

\bibitem{BL}  Brezhnev, Yu.V. and Leble, S.B. (2005) On integration of
the closed KdV dressing chain, arXiv:math-ph/0502052.

\bibitem{B} Bulgac A. (2002), Dilute quantum droplets.  Phys. Rev. Lett. 89, 050402
; Bordag, M. (2004) Nonsmooth backgrounds in quantum field theory.
Physical Review D 70(4)

\end{thebibliography}
\end{document}